\documentstyle[editedvolume,numreferences,psfig]{crckapb}

\begin{opening}
\title{QUANTUM KINETIC THEORY OF TRAPPED ATOMIC GASES}

\author{H.T.C. Stoof}
\institute{Institute for Theoretical Physics, University of Utrecht \\
           Princetonplein 5, 3584 CC Utrecht, The Netherlands}

\end{opening}

\begin{document}

\begin{abstract}
We present a general framework in which we can accurately describe the non-equilibrium dynamics of trapped atomic gases. This is achieved by deriving a single Fokker-Planck equation for the gas. In this way we are able to discuss not only the dynamics of an interacting gas above and below the critical temperature at which the gas becomes superfluid, but also during the phase transition itself. The last topic cannot be studied on the basis of the usual mean-field theory and was the main motivation for our work. To show, however, that the Fokker-Planck equation is not only of interest for recent experiments on the dynamics of Bose-Einstein condensation, we also indicate how it can, for instance, be applied to the study of the collective modes of a condensed Bose gas.\footnotemark[1]
\end{abstract}

\footnotetext[1]{These are the notes for a lecture delivered in July 1998 at the NATO ASI on Dynamics in Leiden, The Netherlands. It is a contribution to
{\it Dynamics: Models and Kinetic Methods for Non-Equilibrium Many-Body Systems}
edited by John Karkheck and to be published by Kluwer academic publishers b.v.}

\section{Introduction}
The most important reason for the present interest in Bose condensed atomic gases, is the possibility to study in detail the dynamics of a superfluid system in this case. In particular, it is possible to compare {\it ab initio} many-body theories for the non-equilibrium dynamics directly to experiment, which is for instance not possible for liquid helium. Two important issues that are of interest to us here are the dynamics of condensate formation and the eigenfrequencies of the collective modes of the Bose condensed gas cloud. In both of these problems it is important to realize that the gas consists of two components, which, due to the harmonic confinement of the gas, are roughly speaking also spatially separated. More precisely, the density profile of the gas can be viewed as a relatively narrow condensate peak on top of a broad thermal background. 

Theoretically, we anticipate that the noncondensate or thermal cloud behaves similarly as a gas above the critical temperature for Bose-Einstein condensation. Its dynamics is therefore accurately described by an appropriate kinetic equation. However, the condensate is a macroscopic quantum object and the dynamics of the condensate must therefore be determined by an equation for its wavefunction. On the basis of these arguments we see that to describe the coupled dynamics of the thermal and condensate clouds, we need a quantum kinetic theory that is capable of simultaneously treating both the incoherent as well as the coherent processes taking place in the gas. It is the main purpose of this lecture to explain as simple as possible how such a quantum kinetic theory can be derived from first principles \cite{crispin1}. Moreover, as an application and illustration of the theory, we also present the first results on the formation of the condensate and the collective modes that we have recently obtained. 

It should be noted that in this lecture we use only physical arguments to explain and motivate the final structure of our quantum kinetic theory. For a detailed derivation by means of field-theoretical methods we refer to the existing literature \cite{henk1,henk2}. In addition, we mainly restrict ourselves here to a discussion of the so-called weak-coupling limit in which interactions have only a relatively small, but nevertheless crucial, effect on the behavior of the gas. This restriction is again made for clarity reasons only, because in the weak-coupling limit the theory is most transparent and
therefore most easily understood. Moreover, once this limit is well understood, it is in principle straightforward to generalize and treat also the strong-coupling limit. However, before we can start to consider the effect of interactions, we first need to reformulate the theory of the ideal Bose gas in a somewhat unusual way, that nevertheless turns out to be very convenient for our purposes. 

\section{Ideal Bose gas}
Let us therefore consider an ideal Bose gas in an external trapping potential 
$V^{\rm trap}({\bf x})$ with one-particle energy levels $\epsilon_{\alpha}$ and corresponding wavefunctions $\chi_{\alpha}({\bf x})$, that can be found from the Schr\"odinger equation
\begin{equation}
\left\{ - \frac{\hbar^2 \mbox{\boldmath $\nabla$}^2}{2m} + V^{\rm trap}({\bf x})
  \right\} \chi_{\alpha}({\bf x}) =  \epsilon_{\alpha} \chi_{\alpha}({\bf x})~.
\end{equation}
Using the methods of second quantization, the non-equilibrium dynamics of the gas is now in the Heisenberg picture fully determined by the initial density matrix $\hat{\rho}(t_0)$ and the hamiltonian
\begin{equation}
\hat{H} = \sum_{\alpha} \epsilon_{\alpha} \hat{N}_{\alpha}(t)
= \sum_{\alpha} \epsilon_{\alpha} \hat{\psi}^{\dagger}_{\alpha}(t)
                                          \hat{\psi}_{\alpha}(t) ~,
\end{equation}
where $\hat{\psi}^{\dagger}_{\alpha}(t)$ and $\hat{\psi}_{\alpha}(t)$ create and annihilate at time $t$ a particle in the state $\chi_{\alpha}({\bf x})$, respectively.

If we would follow the usual treatment of an ideal Bose gas, we would at this point introduce the basis $|\{N_{\alpha}\};t\rangle$ in which the occupation numbers of all the one-particle state $\chi_{\alpha}({\bf x})$ are specified, and proceed to calculate the probability distribution
\begin{equation}
P(\{N_{\alpha}\};t) = {\rm Tr} \left[ \hat{\rho}(t_0)^{^{^{^{}}}} 
                             |\{N_{\alpha}\};t\rangle \langle\{N_{\alpha}\};t| 
                           \right]~.
\end{equation}
For an ideal gas the hamiltonian commutes with the number operators $\hat{N}_{\alpha}(t)$ and the above probability distribution is in fact independent of time. In particular, this implies that the average occupation numbers $\langle \hat{N}_{\alpha}(t) \rangle \equiv 
               {\rm Tr} \left[ \hat{\rho}(t_0) \hat{N}_{\alpha}(t) \right]$ are constant, which physically makes sense because without any interactions there is no way in which the particles can scatter from one state to another.  

As mentioned in the introduction, we are not only interested in the occupation numbers of the gas, but also in the condensate wavefunction. Therefore we do not want to use as a basis the eigenstates of the number operators $\hat{N}_{\alpha}(t)$, but instead the eigenstates of the annihilation operators
$\hat{\psi}_{\alpha}(t)$. More precisely, we introduce the so-called coherent states $|\phi;t \rangle$, which are (properly normalized) eigenstates of the field operator 
\begin{equation}
\hat{\psi}({\bf x},t) 
             = \sum_{\alpha} \chi_{\alpha}({\bf x}) \hat{\psi}_{\alpha}(t)
\end{equation}
with complex eigenvalue $\phi({\bf x})$ \cite{NO}, and consider the corresponding probability distribution
\begin{equation}
P[\phi^*,\phi;t] = 
   {\rm Tr} \left[ \hat{\rho}(t_0)^{^{^{^{}}}} 
              |\phi;t\rangle \langle\phi;t| \right]~.
\end{equation}

We now need to determine the equation of motion, i.e., the appropriate Fokker-Planck equation, for this probability distribution. This can be achieved most easily as follows. We know that by definition the annihilation operators $\hat{\psi}_{\alpha}(t)$ obey the Heisenberg equation
\begin{equation}
i\hbar \frac{\partial \hat{\psi}_{\alpha}(t)}{\partial t}
                    = \left[ \hat{\psi}_{\alpha}(t), \hat{H} \right]_- 
                    = \epsilon_{\alpha} \hat{\psi}_{\alpha}(t)~.
\end{equation}
Since we have used the eigenstates of the annihilation operators to define the probability distribution $P[\phi^*,\phi;t]$, we also know that
\begin{equation}
\langle \hat{\psi}_{\alpha}(t) \rangle =
  \int d[\phi^*]d[\phi]~ \phi_{\alpha} P[\phi^*,\phi;t] 
  \equiv \langle \phi_{\alpha} \rangle(t)~,
\end{equation}
where $\int d[\phi^*]d[\phi]$ denotes the (functional) integral over the complex functions $\phi({\bf x})$. Combining the last two equation we thus find that
\begin{equation}
i\hbar \frac{\partial}{\partial t} \langle \phi_{\alpha} \rangle(t)
                    = \epsilon_{\alpha} \langle \phi_{\alpha} \rangle(t)~.
\end{equation}
Moreover, by considering the Heisenberg equation for the creation operators $\hat{\psi}^{\dagger}_{\alpha}(t)$ we also obtain
\begin{equation}
i\hbar \frac{\partial}{\partial t} \langle \phi^*_{\alpha} \rangle(t)
                    = - \epsilon_{\alpha} \langle \phi^*_{\alpha} \rangle(t)~.
\end{equation}

In this manner we have thus been able to derive the equation of motion for the first moments of the probability distribution. However, to arrive at a Fokker-Planck equation for $P[\phi^*,\phi;t]$ we also need to consider the higher moments \cite{nico}, and in our case in particular 
$\langle |\phi_{\alpha}|^2 \rangle(t)$. {\it A priori} we expect this expectation value to be related to the average occupation numbers  
$\langle \hat{N}_{\alpha}(t) \rangle$ and therefore that
\begin{equation}
i\hbar \frac{\partial}{\partial t} \langle |\phi_{\alpha}|^2 \rangle(t) = 0~.
\end{equation}
Although the latter result is all that we need here, we need lateron also the precise relation between $\langle |\phi_{\alpha}|^2 \rangle(t)$ and the average occupation numbers, which can be shown to be given by
\begin{equation}
\langle |\phi_{\alpha}|^2 \rangle(t) 
         = \langle \hat{N}_{\alpha}(t) \rangle  + \frac{1}{2}~.
\end{equation}
A derivation of this relation is complicated by the fact that the creation and annihilation operators do not commute at equal times. It can, however, be understood physically from the fact that the second moment of $P[\phi^*,\phi;t]$ should contain both classical as well as quantum fluctuations.

From Eqs. (8), (9) and (10) we conclude that the desired Fokker-Planck equation reads \cite{CD} 
\begin{eqnarray}
i\hbar \frac{\partial}{\partial t} P[\phi^*,\phi;t] =
 &-& \sum_{\alpha} \frac{\partial}{\partial \phi_{\alpha}} 
       \left( \epsilon_{\alpha} \phi_{\alpha} \right) P[\phi^*,\phi;t]
                                                                \nonumber \\
 &+& \sum_{\alpha} \frac{\partial}{\partial \phi^*_{\alpha}} 
       \left( \epsilon_{\alpha} \phi^*_{\alpha} \right) P[\phi^*,\phi;t]~.
\end{eqnarray}
It thus contains in the right-hand side only `streaming' terms and no `diffusion' term. As a result there is no unique equilibrium and in fact any function of $|\phi_{\alpha}|^2$ is a stationary solution. Again, this makes sense physically, because without interactions there is no way in which the occupation numbers can relax to an equilibrium Bose-Einstein distribution. To include such relaxation into our discussion, we therefore now bring our ideal Bose gas into contact with a thermal reservoir. 

\section{Ideal Bose gas in contact with a reservoir}
As our reservoir we take an ideal Bose gas in a box with volume $V$. The reservoir is assumed to be sufficiently large so that it can be treated in the thermodynamic limit. Moreover, it is in equilibrium with a temperature $T$ and a chemical potential $\mu$. The states in this box are labeled by the momentum $\hbar{\bf k}$ and equal to 
$\chi_{\bf k}({\bf x}) = e^{i {\bf k} \cdot {\bf x} }/\sqrt{V}$. They have an 
energy $\epsilon({\bf k}) = \hbar^2 {\bf k}^2/2m$. 
Finally, the reservoir is thought to be in contact with the trap discussed above, by means of the tunnel hamiltonian
\begin{equation}
\hat{H}^{\rm int} = \frac{1}{\sqrt{V}}
  \sum_{\alpha} \sum_{\bf k}~
  \left( t_{\alpha}({\bf k}) 
           \hat{\psi}_{\alpha}(t) \hat{\psi}^{\dagger}_{\bf k}(t)
       + t^*_{\alpha}({\bf k}) 
           \hat{\psi}_{\bf k}(t) \hat{\psi}^{\dagger}_{\alpha}(t) \right)~.
\end{equation}
Here $t_{\alpha}({\bf k})$ are complex tunneling matrix elements that for 
simplicity are assumed to be almost constant for momenta $\hbar k$ smaller that 
a cutoff $\hbar k_c$ but to vanish rapidly for momenta larger than this 
ultraviolet cutoff. Moreover, we consider here only the low-temperature regime 
in which the thermal de Broglie wavelength $\Lambda = (2\pi\hbar^2/mk_BT)^{1/2}$ 
of the particles obeys $\Lambda \gg 1/k_c$, since this is the most appropriate 
limit for realistic atomic gases.

Due to this interaction the particles can tunnel back and forth from the trap to the reservoir, which results both in a shift in the energy as well as a finite lifetime of the state $\chi_{\alpha}({\bf x})$. The energies of the states in the trap therefore become complex and equal to 
$\epsilon_{\alpha} + S_{\alpha} - iR_{\alpha}$, where the real and imaginary contributions to the shift can essentially be found from second-order perturbation theory. Denoting the Cauchy principle value part of an integral by ${\cal P}$, they obey
\begin{equation}
\label{shift}
S_{\alpha}
  = \int \frac{d{\bf k}}{(2\pi)^3}~
         t^*_{\alpha}({\bf k}) 
           \frac{{\cal P}}{\epsilon_{\alpha} + S_{\alpha} - \epsilon({\bf k})} 
         t_{\alpha}({\bf k})
\end{equation}
and
\begin{equation}
\label{decay}
R_{\alpha}
  = \pi \int \frac{d{\bf k}}{(2\pi)^3}~
        \delta( \epsilon_{\alpha} + S_{\alpha} - \epsilon({\bf k}) )         
                                          |t_{\alpha}({\bf k})|^2~,
\end{equation} 
respectively.

Introducing the retarded and advanced selfenergies 
$\hbar\Sigma^{(\pm)}_{\alpha} = S_{\alpha} \mp i R_{\alpha}$, we conclude from the above that Eqs. (8) and (9) now become
\begin{equation}
i\hbar \frac{\partial}{\partial t} \langle \phi_{\alpha} \rangle(t)
   = \left( \epsilon_{\alpha} + \hbar\Sigma^{(+)}_{\alpha} - \mu \right) 
                                        \langle \phi_{\alpha} \rangle(t)
\end{equation}
and
\begin{equation}
i\hbar \frac{\partial}{\partial t} \langle \phi^*_{\alpha} \rangle(t)
   = - \left( \epsilon_{\alpha} + \hbar\Sigma^{(-)}_{\alpha} - \mu \right) 
                                        \langle \phi^*_{\alpha} \rangle(t)~,
\end{equation}
where we have also measured our energies relative to the chemical potential.

Next, we need to consider the fluctuations, i.e., the generalization of Eq. (10). This turns out to be given by
\begin{equation}
i\hbar \frac{\partial}{\partial t} \langle |\phi_{\alpha}|^2 \rangle(t) =
  - 2iR_{\alpha} \langle |\phi_{\alpha}|^2 \rangle(t) 
  - \frac{1}{2} \hbar\Sigma^K_{\alpha}~,
\end{equation}
with the so-called Keldysh selfenergie equal to
\begin{equation}
\hbar\Sigma^K_{\alpha} = -2iR_{\alpha} \left( 1 + 2N^{\rm eq}_{\alpha} \right)
\end{equation}
and $N^{\rm eq}_{\alpha} 
       = 1/(e^{(\epsilon_{\alpha} + S_{\alpha} - \mu)/k_BT}-1)$
the equilibrium Bose-Einstein distribution function. How can this result be understood? The first term in the right-hand side of Eq. (18) follows simply from the fact that if we neglect correlations 
$\langle |\phi_{\alpha}|^2 \rangle(t)$ is equal to 
$|\langle \phi_{\alpha} \rangle(t)|^2$. Furthermore, the second term in the right-hand side of Eq. (18) guarantees that if we make use of the relation between $\langle |\phi_{\alpha}|^2 \rangle(t)$ and the average occupation numbers  
$N_{\alpha}(t) \equiv \langle \hat{N}_{\alpha}(t) \rangle$, we recover exactly the Boltzmann equation
\begin{equation}
\frac{\partial}{\partial t} N_{\alpha}(t) 
  = - \Gamma_{\alpha} N_{\alpha}(t) + \Gamma_{\alpha} N^{\rm eq}_{\alpha} 
\end{equation}
with the correct transition rates $\Gamma_{\alpha} = 2R_{\alpha}/\hbar$ expected from Fermi's Golden Rule.

In a similar manner as for the isolated case, we now conclude from Eqs. (16), (17) and (18) that the Fokker-Planck equation for our trapped Bose gas becomes
\begin{eqnarray}
i\hbar \frac{\partial}{\partial t} P[\phi^*,\phi;t] = 
  &-& \sum_{\alpha} \frac{\partial}{\partial \phi_{\alpha}} 
       \left( \epsilon_{\alpha} + \hbar\Sigma^{(+)}_{\alpha} - \mu \right)
                                                \phi_{\alpha} P[\phi^*,\phi;t]
                                                \nonumber \\ 
  &+& \sum_{\alpha} \frac{\partial}{\partial \phi^*_{\alpha}} 
       \left( \epsilon_{\alpha} + \hbar\Sigma^{(-)}_{\alpha} - \mu \right) 
                                              \phi^*_{\alpha} P[\phi^*,\phi;t]
                                              \nonumber \\
  &-& \frac{1}{2} \sum_{\alpha} 
       \frac{\partial^2}{\partial \phi_{\alpha} \partial \phi^*_{\alpha}}
       \hbar\Sigma^K_{\alpha} P[\phi^*,\phi;t]~.
\end{eqnarray}
It is interesting to note that this Fokker-Planck equation is equivalent to the
Langevin equation
\begin{equation}
i\hbar \frac{\partial}{\partial t} \phi_{\alpha}(t)
   - \left( \epsilon_{\alpha} + \hbar\Sigma^{(+)}_{\alpha} - \mu \right) 
                   \phi_{\alpha}(t)  = \eta_{\alpha}(t)~,
\end{equation}
where the gaussian noise obeys
\begin{equation}
\langle \eta^*_{\alpha}(t) \eta_{\alpha'}(t') \rangle
  = \frac{i\hbar^2}{2} \hbar\Sigma^K_{\alpha} \delta_{\alpha,\alpha'}
                                              \delta(t-t')
\end{equation}
and the fluctuation-dissipation theorem reads
\begin{equation}
\hbar\Sigma^K_{\alpha} = \left( 1 + 2N^{\rm eq}_{\alpha} \right) 
   \left( \hbar\Sigma^{(+)}_{\alpha} - \hbar\Sigma^{(-)}_{\alpha} \right)~.
\end{equation}

Clearly, due to the fluctuation-dissipation theorem the probability distribution $P[\phi^*,\phi;t]$ for the gas in the trap now relaxes to the correct equilibrium state    
\begin{equation}
P[\phi^*,\phi;\infty] = \prod_{\alpha} \frac{1}{N^{\rm eq}_{\alpha} + 1/2}
     \exp \left\{ - \frac{1}{N^{\rm eq}_{\alpha} +1/2} |\phi_{\alpha}|^2
          \right\}~.
\end{equation}
More important for our purposes is, however, that the Fokker-Planck equation in Eq. (21) describes simultaneously both the incoherent (kinetic) as well as the coherent dynamics in the gas, since it incorporates the equations of motion for both $\langle \hat{N}_{\alpha}(t) \rangle$ and 
$\langle \hat{\psi}_{\alpha}(t) \rangle$, respectively. As mentioned previously, this is precisely what is needed for an accurate treatment of non-equilibrium phenomena in Bose condensed atomic gases.

\section{Condensate formation in an interacting Bose gas}
We are now in a position to discuss an interacting Bose gas, because in an interacting Bose gas, the gas is roughly speaking its own thermal reservoir. The Fokker-Planck equation therefore turns out to be quite similar to the one presented in Eq. (21). To be more precise, however, we should mention that we aim in this section to describe the formation of a condensate in an interacting Bose gas under the conditions that have recently been realized in experiments with atomic $^{87}$Rb \cite{JILA}, $^7$Li \cite{Rice} and $^{23}$Na \cite{MIT} gases. In these experiments the gas is cooled by means of evaporative cooling. Numerical solutions of the Boltzmann equation have shown that during evaporative cooling the energy distribution function is well described by an equilibrium distribution with time dependent temperature $T(t)$ and chemical potential $\mu(t)$, that is truncated at high energies due to the evaporation of the highest energetic atoms from the trap \cite{jom}. Moreover, in the experiments of interest the densities just above the critical temperature are essentially always such that the gas is in the weak-coupling limit, which implies in this context that the average interaction energy per atom is always much less than the energy splitting of the one-particle states in the harmonic trapping potential.

Keeping the above remarks in mind, we find that near the critical temperature the non-equilibrium properties of the gas are in an excellent approximation described by a nonlinear Fokker-Planck equation with time-dependent selfenergies $\hbar\Sigma^{(\pm),K}_{\alpha}(t)$ and effective interaction matrix elements $V^{(\pm),K}_{\alpha,\beta;\alpha',\beta'}(t)$, for which in the so-called ladder approximation explicite expressions can be derived in terms of the average occupation numbers $N_{\alpha}(t)$ in the gas \cite{henk2}. In full detail it reads 
\begin{eqnarray}
i\hbar \frac{\partial}{\partial t} P[\phi^*,\phi;t] = && \nonumber \\
&& \hspace{-1.3in}
 - \sum_{\alpha}
   \frac{\partial}{\partial \phi_{\alpha}}
       \left\{ \left( \epsilon_{\alpha} + \hbar\Sigma^{(+)}_{\alpha} - \mu
               \right) \phi_{\alpha}
            + \sum_{\alpha',\beta,\beta'}
               V^{(+)}_{\alpha,\beta;\alpha',\beta'}
                   \phi^*_{\beta}
                   \phi_{\beta'} \phi_{\alpha'}
       \right\} P[\phi^*,\phi;t]                         \nonumber \\
&& \hspace{-1.3in}
+ \sum_{\alpha}
  \frac{\partial}{\partial \phi^*_{\alpha}}
       \left\{ \left( \epsilon_{\alpha} + \hbar\Sigma^{(-)}_{\alpha} - \mu
               \right) \phi^*_{\alpha}
            + \sum_{\alpha',\beta,\beta'}
               V^{(-)}_{\alpha',\beta';\alpha,\beta}
                   \phi^*_{\alpha'} \phi^*_{\beta'}
                   \phi_{\beta}
       \right\} P[\phi^*,\phi;t]                         \nonumber \\
&& \hspace{-1.3in}
- \frac{1}{2} \sum_{\alpha,\alpha'}
  \frac{\partial^2}{\partial \phi_{\alpha}
                    \partial \phi^*_{\alpha'}}
       \left\{ \hbar\Sigma^K_{\alpha} \delta_{\alpha,\alpha'}
            + \sum_{\beta,\beta'}
               V^K_{\alpha,\beta;\alpha',\beta'}
                   \phi^*_{\beta}\phi_{\beta'}
       \right\} P[\phi^*,\phi;t]~.
\end{eqnarray} 

To understand how this equation describes the formation of the condensate, we make use of the fact that in the weak-coupling limit it is appropriate to solve the Fokker-Planck equation with the (Hartree-Fock) {\it ansatz}
\begin{equation}
P[\phi^*,\phi;t] = P_0[\phi^*_g,\phi_g;t] P_1[\phi'^*,\phi';t]~,
\end{equation}
where the complex function
$\phi'({\bf x}) = \sum_{\alpha \neq g} \chi_{\alpha}({\bf x}) \phi_{\alpha}$
is associated with all the one-particle states except the groundstate $\chi_g({\bf x})$. Substituting the above {\it ansatz} in the Fokker-Planck equation, we obtain the following results. First, the dynamics of the noncondensed cloud is determined by the quantum Boltzmann equation
\begin{equation}
\frac{\partial}{\partial t} N_{\alpha}(t) 
  = - \Gamma^{\rm out}_{\alpha} N_{\alpha}(t) + 
      \Gamma^{\rm in}_{\alpha} (1 + N_{\alpha})~, 
\end{equation}
with $\alpha \neq g$ and the scattering rates $\Gamma^{\rm out,in}(t)$ of a similar form as in the Uehling-Uhlenbeck equation but with a cross-section which is proportional to $|V^{(+)}_{\alpha',\beta';\alpha,\beta}(t)|^2$. Second, this kinetic equation is coupled to a Fokker-Planck equation for the condensate, i.e., to
\begin{eqnarray}
i\hbar \frac{\partial}{\partial t} P_0[\phi^*_g,\phi_g;t] = 
 &-& \frac{\partial}{\partial \phi_g}
       \left( \epsilon_g + \hbar\Sigma^{(+)}_g - \mu
               + V^{(+)}_{g,g;g,g} |\phi_g|^2 \right) \phi_g
       P_0[\phi^*_g,\phi_g;t]                         \nonumber \\
&+& \frac{\partial}{\partial \phi^*_g}
       \left( \epsilon_g + \hbar\Sigma^{(-)}_g - \mu
              + V^{(-)}_{g,g;g,g} |\phi_g|^2 \right) \phi^*_g
       P_0[\phi^*_g,\phi_g;t]                         \nonumber \\
&-& \frac{1}{2} 
  \frac{\partial^2}{\partial \phi_g
                    \partial \phi^*_g}
       \hbar\Sigma^K_g P_0[\phi^*_g,\phi_g;t]~.
\end{eqnarray} 

Due to the fluctuation-dissipation theorem, the probability distribution for the condensate relaxes to the equilibrium solution
\begin{equation}
P_0[\phi^*_g,\phi_g;\infty] \propto \exp \left\{ - \frac{1}{k_BT}
   \left( \epsilon_g + S_g - \mu 
          + \frac{V^{(+)}_{g,g;g,g}}{2} |\phi_g|^2 \right)
                            |\phi_g|^2 \right\}~.  
\end{equation}
The Landau free energy for the condensate order parameter thus equals
\begin{equation}
F_L[\phi^*_g,\phi_g] = \left( \epsilon_g + S_g - \mu \right) |\phi_g|^2
          + \frac{V^{(+)}_{g,g;g,g}}{2} |\phi_g|^4 
\end{equation}
and clearly shows a spontaneous breaking of symmetry if 
$\epsilon_g + S_g - \mu < 0$ and the effective interatomic interaction is repulsive, i.e., $V^{(+)}_{g,g;g,g} > 0$ \cite{henk3}. In our formulation of the problem these quantities are a function of time, whose evolution is essentially determined by the quantum Boltzmann equation in Eq. (28). In particular, for Bose-Einstein condensation to occur, $\epsilon_g + S_g - \mu$ has to change sign during this evolution. As a first rough calculation of the condensate formation, we can assume that this change of sign takes place instantaneously. Introducing for convenience the dimensionless time 
$\tau = t (i\Sigma^K_g/8)(2V^{(+)}_{g,g;g,g}/k_BT)^{1/2}$ and the dimensionless condensate number 
$I = |\phi_g|^2 (2V^{(+)}_{g,g;g,g}/k_BT)^{1/2}$, this assumption leads to a typical evolution for the probability distribution $P_0[I;\tau]$ that is shown in Fig. 1.

\begin{figure}[hb]
\psfig{figure=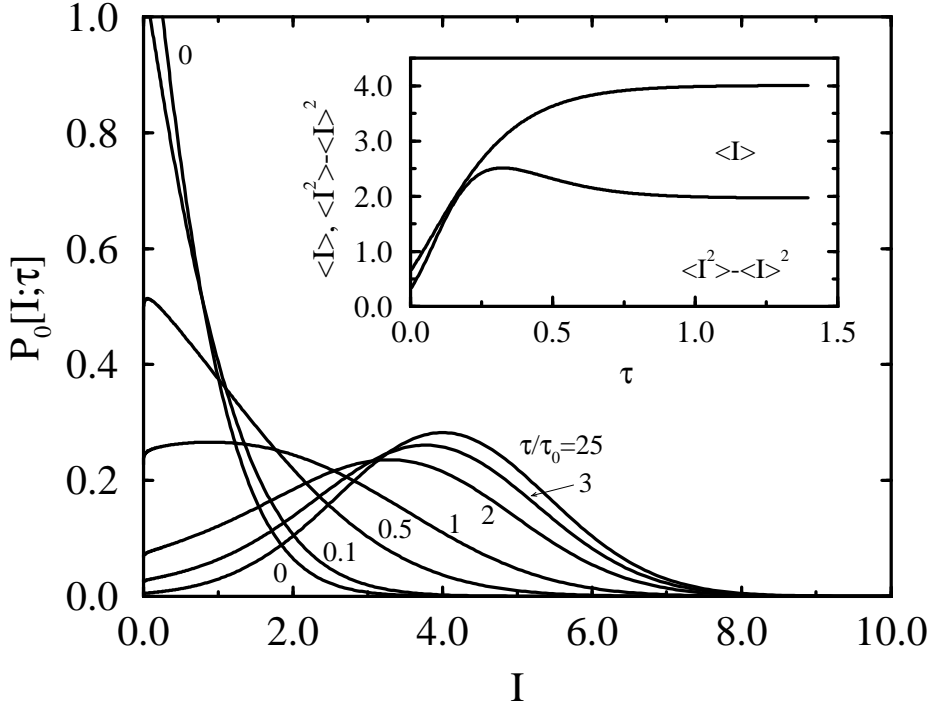}
\caption{Typical evolution of the condensate probability distribution during
         Bose-Einstein condensation. The slowest relaxation rate is 
         $1/\tau_0 \simeq 5.7$. The inset shows the evolution of the
         average condensate number and its fluctuations.
         \label{growth}}
\end{figure}

At this point we should mention that the solution of the condensate Fokker-Planck equation in Eq. (29) with a fixed and negative value of 
$\epsilon_g + S_g - \mu$, is equivalent to the theory recently put forward by Gardiner and coworkers using different methods \cite{crispin2}. It even turns out to agree qualitatively with experiment, although quantitatively some important discrepancies exists \cite{wolfgang}. In our opinion these discrepancies are probably due to the fact that we should really solve the Fokker-Planck equation for the condensate together with the quantum Boltzmann equation for the thermal cloud. Moreover, and additional problem is that the experiments of interest here are not really in the weak-coupling limit, which substantially complicates the theory because more states are needed to describe
the condensate. 

\section{Collective modes}
As another illustration of our general nonequilibrium approach we consider now the collective modes of a Bose condensed gas at such high temperatures that a substantial noncondensate fraction is present in the gas. The experiments that have been performed at these relatively high temperatures, are in the so-called collisionless regime \cite{eric1}. Physically, this implies that the osscilation period of the mode of interest is much shorter that the average time between two collisions of the atoms. Under these conditions, our Fokker-Planck equation for the gas gives, in the Hartree-Fock approximation and after a transformation of Eq. (26) to coordinate space, first of all a collisionless Boltzmann or Vlasov-Landau equation for the long wavelength Wigner distribution $N({\bf x},{\bf k};t)$ of the noncondensed part of the gas. Explicitly, it reads
\begin{eqnarray}
\frac{\partial}{\partial t} N({\bf x},{\bf k};t)  
  &+& \frac{\hbar {\bf k}}{m} \cdot
      \frac{\partial}{\partial {\bf x}} N({\bf x},{\bf k};t) 
                                                     \nonumber \\
  &-& \frac{1}{\hbar}
    \frac{\partial}{\partial {\bf x}} 
    \left( V^{\rm trap}({\bf x}) + 2V^{(+)}n({\bf x},t) \right) \cdot
      \frac{\partial}{\partial {\bf k}} N({\bf x},{\bf k};t) = 0~,~~~~~~
\end{eqnarray}    
where the effective interaction $V^{(+)} = 4\pi\hbar^2 a/m$ can be expressed in the two-body $s$-wave scattering length $a$ and $n({\bf x},t)$ is the total density of the gas. Furthermore, it leads to a nonlinear Schr\"odinger equation for the condensate wavefunction $\langle \phi({\bf x}) \rangle(t)$, i.e., to
\begin{eqnarray}
i\hbar \frac{\partial}{\partial t} \langle \phi({\bf x}) \rangle(t) =  
                                                \hspace*{3.3in} \nonumber \\
   \left\{ - \frac{\hbar^2 \nabla^2}{2m} + V^{\rm trap}({\bf x}) - \mu 
           + V^{(+)}(2n'({\bf x},t) + n_0({\bf x},t)) \right\}
                         \langle \phi({\bf x}) \rangle(t)~, 
\end{eqnarray}
with the condensate density 
$n_0({\bf x},t)) = |\langle \phi({\bf x}) \rangle(t)|^2$ and the noncondensate density 
$n'({\bf x},t) = \int d{\bf k}~ N({\bf x},{\bf k};t)/(2\pi)^3$
of course adding up to the total density of the gas.

A numerical solution of the above coupled equations turns out to be surprisingly difficult. To nevertheless gain insight in the collective modes of the gas, we have therefore recently put forward a variational approach, in which we apply a dynamical scaling {\it ansatz} on the ideal gas results for both the Wigner distribution $N({\bf x},{\bf k};t)$ and the condensate wavefunction 
$\langle \phi({\bf x}) \rangle(t)$. The outcome of this calculation is presented in Fig. 2, where we also make a comparison with experiment which turns out to be quite reasonable in view of the simplicity of our scaling 
{\it ansatz}. The main discrepancies are the two measurements halfway between the $m=0$ in and out-of-phase modes. This discrepancy is, however, presumably due to the fact that in the experiment both modes are excited simultaneously \cite{eric2}.

\begin{figure}[hb]
\psfig{figure=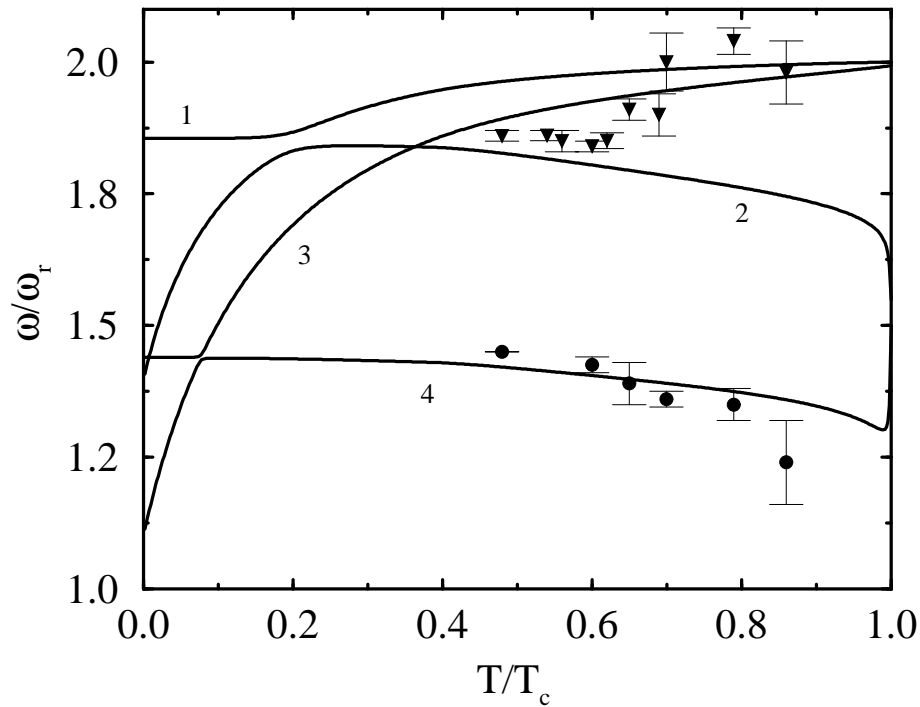}
\caption{Collisionless modes in a Bose condensed $^{87}$Rb gas. Curves 1 and 2 
         correspond to the $m=0$ in and out-of-phase modes, respectively.
         Similarly, curves 3 and 4 give the frequencies of the $m=2$ in and
         out-of-phase modes. The experimental data is also shown. Triangles 
         are for a $m=0$ mode and circles for a $m=2$ mode.
         \label{modes}}
\end{figure} 

\section{Conclusions}
In summary, we have presented a general framework in which we can discuss various non-equilibrium problems in Bose condensed atomic gases. As an example we have presented our first results on the formation of the condensate and the collisionless modes at nonzero temperatures. Other topics that are of interest are, for example, the condensate formation in a Bose gas with attractive interactions, the dynamics of vortices and spin domain walls, the collective modes in the hydrodynamic regime, the damping of collective modes, and last but not least the non-equilibrium dynamics of trapped atomic Fermi gases. In view of the many topics that remain to be explored, we hope that the present lecture may motivate some of the participants of this summerschool to enter into this, in our view, very exciting area of physics. 

\section*{Acknowledgments}
It is a great pleasure to thank Marianne Houbiers and Michiel Bijlsma for their important contributions to the topics discussed in this paper.

\end{document}